\documentclass[a4paper, english,11pt]{article}
\usepackage[T1]{fontenc}
\usepackage{babel}
\usepackage{textcomp}
\usepackage{amsmath}
\usepackage{amsmath}
\usepackage{setspace}
\usepackage{amssymb}
\usepackage{lscape}
\usepackage{caption}
\usepackage{subfig}
\usepackage{colortbl}
\usepackage{eucal}
\usepackage{amsfonts}
\usepackage{listings}
\usepackage[]{sidecap}
\usepackage{hyperref}
\usepackage{authblk}
\usepackage[a4paper,top=1.5in,bottom=1.5in,left=1in,right=1in, bindingoffset=5mm]{geometry} 
\usepackage{pdfpages}

\begin{document}
\title{The efficiency of Anderson-Darling test with limited sample size: an application to Backtesting Counterparty Credit Risk internal model\footnote{The views, thoughts and opinions expressed in this paper are those of the authors in their individual capacity and should not be attributed to UniCredit S.p.A. or to the authors as representatives or employees of UniCredit S.p.A.}}

\author[1,2]{Matteo Formenti\thanks{Corrisponding Author, \href{mailto:mformenti@liuc.it}{mformenti@liuc.it}}}
\author[1,3]{Luca Spadafora}
\author[1]{Marcello Terraneo}
\author[1]{Fabio Ramponi}
\affil[1]{UniCredit S.p.A., Piazza Gae Aulenti 3, 20154 Milan, Italy}
\affil[2]{Universit\`a Carlo Cattaneo - LIUC,  Scuola di Economia e Management, C.so Matteotti, 22 - 21053 Castellanza (VA)} 
\affil[3]{Universit\`a Cattolica del Sacro Cuore, Faculty of Mathematical, Physical and Natural Sciences, via dei Musei, 41 - 25121 Brescia} 
\date{\today}
\maketitle
\begin{center}
\vspace{-1cm}
\large(Pre-Print Version)
\end{center}
\bigskip
\normalsize
\begin{abstract}
This work presents a theoretical and empirical evaluation of Anderson-Darling test when the sample size is limited. The test can be applied in order to backtest the risk factors dynamics in the context of Counterparty Credit Risk modelling. We show the limits of such test when backtesting the distributions of an interest rate model over long time horizons and  we propose a modified version of the test that is able to detect more efficiently an underestimation of the model's volatility. Finally we provide an empirical application. 

\end{abstract}
  \thispagestyle{empty}
\newpage{}
\pagenumbering{arabic}

\section*{Introduction}
\label{Backtesting}
Backtesting is defined as "the quantitative comparison of a model's forecasts against realized values" (BCBS, \cite{bkt}). In Counterparty Credit Risk (CCR), the model forecasts regard the estimates of interest rates, credit spreads, equity or commodity values, that are the underlying risk factors driving the mark-to-market of OTC derivatives, up to the longest maturity of the contracts.
As remarked by BCBS \cite{bkt}, banks choose their own best and appropriate method to aggregate, and then validate, the overall quality of the model forecasts. This can be done through a synthetic value, such as the outcome of a statistical test, having the goal to detect weakness of model forecasts. We remark that such forecasts are computed up to the longest maturity and directly affect the exposure towards a counterparty.
So backtesting is one of the instruments through which Risk Management assesses the forecast of counterparty exposure, and indeed the bank's risk weighted asset value. A failure in that test advocates a model change such as a different model parametrization or even a change in model assumptions (e.g. log-normal or t-distribution).

Banks, having a CCR internal model, compute backtesting: (i) on risk-factors level with the aim to validate the properties of the stochastic process used to simulate interest rates, credit spreads, forex and equities; (ii) on trades level, such as plain vanilla or exotic options aiming at validate the single deal exposure; (iii) on counterparty level to validate the soundness of the estimated exposures. We remark that risk managers are interested in backtesting all the forecast distribution shape for a given risk factor, in order to detect an underestimation of risk (i.e., variance) that may lead to an underestimation of the Regulatory Capital measure (RWA) or the managerial measures (Expected Positive Exposure, Potential Future Exposure).

In this paper we study from an empirical and theoretical point of view the statistical properties of the Anderson-Darling (AD) test (\cite{AD}, \cite{AD2010}) used to backtest risk factors, in the particular case of limited sample size. In fact, as AD test has been widely used in empirical literature, from biology to sociology, due to its well known statistical properties for large sample (see \cite{Daniel} and \cite{Gibbons}), its robustness when sample is limited has never been studied, to the best of our knowledge, in the case of CCR modeling. 
On the other hand, regulation asks banks to verify the modeling choice through a backtesting program (CRR art. 293-b), to use at least three years of historical data for model estimation purposes (CRR art. 292-2) and to have procedures in place to identify and control the risks for counterparties where the exposure rises beyond the one-year horizon (CRR, art.289-6). With this respect, we observe that from a purely statistical point of view, the use of overlapping time widows to verify model performance cannot be considered a significant improvement for what concern the reduction of statistical uncertainty of the forecasted variables. In fact, it can be shown that in our context, if the random variables are i.i.d., the additional information included in a statistical estimator based on overlapping time windows, compared with the one related to a non-overlapping estimator is not enough to reduce significantly the related statistical uncertainty. As a consequence, a backtesting methodology based on overlapping time windows would face similar statistical issues related to the small sample size as the ones related to a non-overlapping methodology.
Furthermore, the limited sample size is also an unavoidable condition linked to the limited length of available market data history (e.g. USD/EUR started in 1999). For these reasons, we focus on the AD statistical properties mainly when the backtesting dataset is made by 5-10 observations. 

It is important to remark that for risk management purposes, and in particular for CCR backtesting validation, test should be able to detect a volatility underestimation that is more dangerous from a risk management point of view than volatility overestimation. This means that test rejection power should be higher when the forecasted distributions has a smaller variance than the empirical one. In addition, this property should also hold in case of limited sample size in order to deal with practical situations where the dataset is typically small. For these reasons, we propose a modified version of AD test that may help risk manager to detect easily the underestimation of volatility.

We highlight that uniformity tests, such as the AD, Kolmogorov-Smirnov (KS)~\cite{KS}, Jarque-Brera~\cite{JB} or Cramer-von Mises (CM)~\cite{CvM}, help to statistically validate the model forecasting values because, at each backtesting date, we can map the realized value to the corresponding p-value of the forecasted distribution. In particular, for a given risk factor (e.g. interest rate, foreign exchange, commodity) $r$, backtesting date $t$ and time horizon $s$, the p-value $F(r)$ corresponding to the realization $r(t+s)$ is computed according to the following algorithm:
\[
\begin{cases}
F(r)=\frac{1}{N+2} &  \text{for } r(t+s) < \hat{r}^{(1)} \\
\\
F(r)=\frac{i+1}{N+2} &  \text{for } \hat{r}^{(i)} < r(t+s) < \hat{r}^{(i+1)} \\
\\
F(r)=\frac{N+1}{N+2} &  \text{for } r(t+s) > \hat{r}^{(N)} \\
\end{cases}
\]
where $\hat{r}^{(i)}$ represents the $i-$forecasted value, and $N$ the total number of forecasted values. As a consequence, the collection of the ordered mapped values should be uniformly distributed if the model is perfectly matching the realized values.\\
An example of the application of this tests in the Counterparty Credit Risk backtesting framework can be found in~\cite{Anfuso}. 

The structure of the paper is as follows. In section \ref{AD-sect} we briefly summarize the AD test and its reduced efficiency when the sample size is limited. Section \ref{AD-asym} proposes a modified version of AD test in order to detect faster an underestimation of the volatility. Then in section \ref{NumResults} we compute a numerical exercise that use real data to generate 'fictitious' time series in order to compare  the AD test, and our modified version, with respect to KS test. Last, in section \ref{Empirical Results} we show as a case study the backtesting of the Black-Karasinski model applied to the Euribor 6 Month. Section \ref{conclusion} summarizes our results.

\section{Anderson-Darling Test}
\label{AD-sect}
Anderson and Darling \cite{AD} designed a statistical test in order to determine whether a given sequence of random variables $X=\{x_1,\dots,x_n\}$ comes from a theoretical cumulative distribution function (CDF) $F(x)$. The null hypothesis $H_0$ is that the data follow $F(x)$, so this test should be used to prove that the data do not follow $F(x)$, given a confidence level.
The AD test is based on the estimation of the following random variable:
\begin{equation}
\label{AD-integral}
W^{2}= \int_{-\infty}^{+\infty} \frac{[ F_{n}(x) - F(x) ]^{2}}{\left( \frac{F(x) ( 1-F(x))}{n} \right) }\mathrm{d}F(x)
\end{equation}
where $F(x)$ is the target CDF and $F_n(x)$ is the empirical distribution derived from the data. The numerator of Eq.~\eqref{AD-integral} represents the distance of the theoretical distribution from the empirical one, while the denominator represents the variance of the empirical estimation of $F(x)$ when the central limit theorem holds, i.e. when $n$ is large enough. In other words, Eq.~\eqref{AD-integral} represents the average of the squared errors between the two distributions (theoretical and empirical) weighted by the implicit uncertainty due to the estimation method of the empirical CDF (order statistics).  As the CDF of a random variable is always distributed uniformly between zero and one (i.e. $F(x)\in U(0,1)$), $W^2$ is a function of uniformly distributed random variables when $H_0$ and the central limit theorem hold. In particular, it does not depend on the distribution $F(x)$.

In this context, we observe that when the variance of the uniform distribution ($F(x) ( 1-F(x))/n$) is close to zero, i.e. for rare events, $F(x) \sim 0$ or $F(x) \sim 1$, the squared error is magnified by the small denominator; in this sense we consider the AD test as more sensitive with respect to the tails of the distribution. On the other hand, we point out that, if the number of observations used to perform the AD test is low, the denominator of the integral in Eq.~\ref{AD-integral} is large, i.e. a large variance is associated to the difference between the theoretical and the empirical distribution; as a consequence also a large difference between the distributions would fall inside the variance amplitude and the AD test will not be able to reject $H_0$ as the uncertainty in the measurement will be too large to bring to any conclusion. So, in order to reject $H_0$, the difference between the theoretical CDF and the empirical one has to be larger than their statistical uncertainty. 
Eq.~\eqref{AD-integral} can be also expressed as:
\begin{equation}\label{AD-SS}
W^{2} = -n - \sum_{k=1}^{n} \frac{2k - 1}{n} \textrm{ln}(F(x_k)) + \textrm{ln}(1-F(x_{n+1-k}))
\end{equation}
where $x_i \in X$  are the empirical data ordered from the smallest to the largest values and $n$ is the size of the sample (i.e. number of backtesting dates). The empirical distribution of $W^{2}$ was estimated by AD (\cite{AD}) and we report the percentiles of $W^{2}$ distribution in Tab.~(\ref{tab_AD}) with the first line indicating the upper tail probabilities and the second line representing the corresponding percentiles.\\
\begin{table}[h]
\centering
\begin{tabular}{|l| c c c c c c c |}
\hline\hline
Probability & 0.250  &   0.150 & 0.100& 0.050 & 0.010 & 0.005 & 0.001\\
Percentile & 1.248  &   1.610& 1.933 & 2.492  & 3.880 & 4.500 & 6.000\\
\hline 
\hline
\end{tabular}
\caption{Upper tail percentiles for Anderson-Darling $W^{2}$ test.}
\label{tab_AD}
\end{table} 

\subsection{Analytical Result: the Efficiency of AD Test} \label{analytical}
The AD test can be used for CCR backtesting purposes as a tool to verify whether the model forecasted distributions are comparable with the empirical one, for a given confidence level.
In this context, the null hypothesis shall be that the two distributions are equal. So the test gives positive outcome when the null hypothesis is not rejected, meaning the model distribution is not sufficiently different from the empirical one to consider the model as wrong. 

\begin{figure}[h]
\centering
\includegraphics[scale= 0.2]{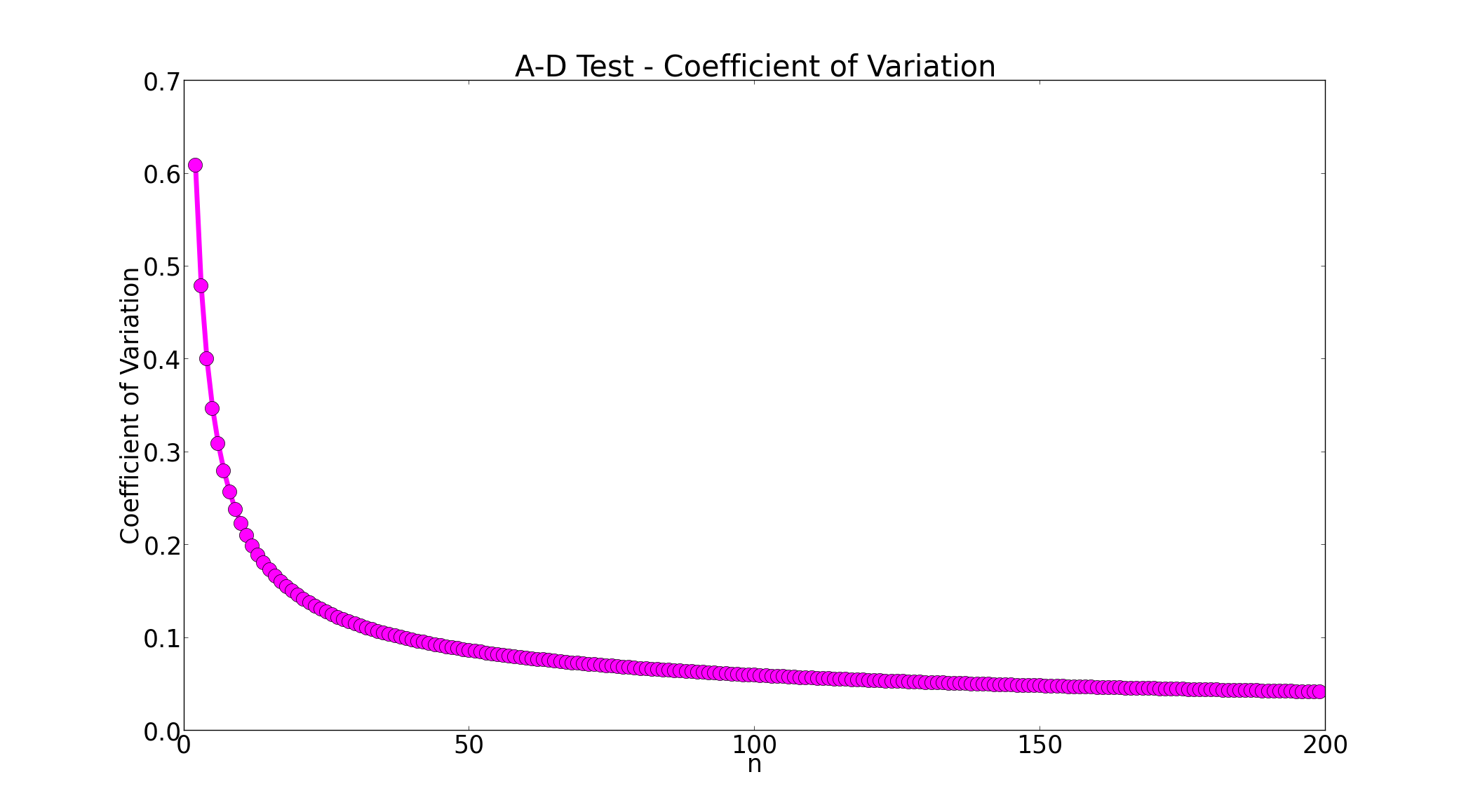}
\caption{Decay of the coefficient of variation as a function of the number of observations}
\label{fig_coefficient_variation}
\end{figure}

Given our backtesting approach, a large uncertainty in the empirical CDF estimation is due to the small number of observations. This negatively affects the rejection rate of our test accepting distributions just because of the limited sample size. As a consequence, we question about the introduction of an efficient measure in such test in order to increase its accuracy when the sample size is small.\\
In general, a measure is considered efficient and faithful if its uncertainty is much smaller than its expected value\footnote{The approach is not consistent if the Expected value is equal to zero}; in our case the statistical uncertainty is related to the empirical estimation of the CDF, while the expected value is the theoretical CDF for each point of the empirical distribution. 
Unfortunately the simple variance estimation for each point of the distribution is not enough as the AD measure requires a sum over all the probabilities $F_n(x_i)$ for $\{i=1,\dots,n\}$, and one has to consider also the whole covariance matrix of the order statistics.
So at each single point the expected value of our empirical estimation is given by $F(x_i) = p_i$, and the covariance structure is given by $p_{\text{min}(i,j)}- p_i p_j$~\cite{Schervish}. We can define the coefficient of variation (CoV) or relative standard deviation, for each point of the distribution as:
\begin{align}\label{CoV}
c &= \frac{\sigma}{\mu} \\
  &= \frac{1}{\sqrt{n}} \frac{\sqrt{\int_{1/n}^{1}\mathrm{min}\left( p,q\right) - p q \mathrm{d}p\mathrm{d}q}}{\int_{1/n}^{1}p \mathrm{d}p}\\
  &=\frac{\sqrt{\frac{1}{3}\left(\frac{1}{n} -1\right)^2\left(\frac{2}{n} + \frac{3}{4}\right) }}{\sqrt{n}\left(\frac{1}{2} - \frac{1}{2 n^2}\right)}\\
  &= \frac{1}{n+1}\sqrt{n + \frac{8}{3}}\\
\end{align}
where $\sigma$ and $\mu$ represent the standard deviation and the expected value of the sum of over estimated probabilities on a range between $1/n$ and $1$, i.e. $\int_{1/n}^{1} F_{n}(x)dF(x)$.
In Fig. \ref{fig_coefficient_variation} we show the decay of the coefficient of variation as a function of the number of observations. In order to obtain a CoV below $10\%$, $n=50$ observations are required. The CoV indicator gives important information about the performance of the AD test when the sample size is small, and it can be used as a warning level when AD test is applied. On the other hand, we remark that CoV is derived assuming the Central Limit Theorem (CLT) holds, whereas this hypothesis does not hold for very small sample sizes. Therefore one should consider additional corrections to Eq. \ref{CoV} that we do not take into consideration for this work.
\\
\section{Asymmetric Extension of AD Test}\label{AD-asym}
In order to detect the underestimation of the actual volatility when the number of observations is small, we derive an extension of the AD test that can be used for risk management purposes. The main idea for AD test extension relies on the observation that, when the sample size is small, it is easier to reject the null hypothesis when the empirical variance of the distribution is larger than the forecasted one;
on the contrary, when the empirical variance is smaller than the forecasted one, many observations fall \textit{inside} the theoretical distribution, so it is more difficult to reject the $H_{0}$ hypothesis (this fact is further discussed with numerical examples in the next sections). Starting from this observation, one could magnify this asymmetric effect defining an AD-Asym function that introduces a more pronounced non linear behaviour when the differences between the theoretical and the empirical CDF are large, as in the case of variance underestimation. 
We stress that the term \emph{asymmetric} refers to the different behavior of the AD test when the variance of the forecasted distribution is over/under estimated and not to the analytical form of $W^2$.
We generalized Eq.~\eqref{AD-integral} as:
\begin{equation}
\label{AD-Asym}
W^{2}_{Asym}= \int_{-\infty}^{+\infty} \frac{[ F_{n}(x) - F(x) ]^{2\beta}}{\left( \frac{F(x) ( 1-F(x))}{n} \right)^\beta } \mathrm{d}F(x)
\end{equation}
where $\beta \ge 1$ is the parameter that controls the amplitude of the asymmetric effect; obviously when $\beta=1$ we recover Eq.~\eqref{AD-integral}. In this paper we focus our attention on the special case $\beta=2$. In this way, the small variance amplitude due to the small $n$ is compensated by the $\beta$ exponent. \\
In order to apply this new asymmetric formulation of the AD-test in practical situations, we have to:
\begin{itemize}
\item estimate the integral in the Eq.~\eqref{AD-Asym};
\item obtain the distribution of the $W^2$ random variables assuming that $H_0$ is true;
\item compare the empirical value of $W^2$ with the theoretical distribution obtained at the previous point and decide if $H_0$ should be rejected at a given confidence level.
\end{itemize}
Once a good estimation of Eq.~\eqref{AD-Asym} is obtained, the second and the third steps can be overcome by a numerical simulation of the $W^2$ r.v. and considering the obtained CDF. On the contrary, some care is required for the integral estimation given the small sample size. In particular, following the lines in~\cite{AD}, we consider a sample $\{x_1,\dots,x_n\}$ observations and we define $x_0 = 0$ and $u = F(x)$ so
\begin{eqnarray}\label{AD-Asym2}
W^{2}_{Asym} &=& \sum_{k=1}^{n}\int_{u_{k-1}}^{u_k} \frac{\left(u - \frac{k-1}{n}\right)^4}{\left(\left(1-u\right)\frac{u}{n}\right)^2} \mathrm{d}u \\
    &=& \sum_{k=1}^{n}- \left(\frac{k-1}{n}\right)^4\frac{1}{u} + 2\left(\frac{k-1}{n}-2\right)\left(\frac{k-1}{n}\right)^3 \mathrm{log}(u) \nonumber\\
    &-& \frac{\left(\frac{k-1}{n} - 1\right)^4}{u - 1} - 2\left(\frac{k-1}{n} +1\right)\left(1 - \frac{k-1}{n}\right)^3 \mathrm{log}(1 - u) + u \Bigg|_{u_{k-1}}^{u_k} \\
    &=& \gamma + \sum_{k=1}^{n-1} \frac{\alpha_1(k)}{u_k} + \alpha_2(k)\mathrm{log}(u_k) + \frac{\alpha_3(k)}{u_{k} - 1} + \alpha_4(k) \mathrm{log}(1 - u_k) \label{AD-Asym3}
\end{eqnarray}
where $u_k = F_n(x_k)$ and $\gamma, \alpha_1(k), \alpha_2(k),\alpha_3(k),\alpha_4(k)$ are functions reported in the Appendix. Eq.~\eqref{AD-Asym3} gives a new AD-Asym indicator that \textit{measures} the differences between empirical and theoretical CDF, emphasising outliers discrepancies. In the following sections, we refer to Eq.~\eqref{AD-Asym3} as AD-Asym indicator in order to compare its performances with respect to the standard AD test. 

\subsection{Testing Anderson-Darling in limited sample size}\label{AD-Asym-test}
In this section we compare the AD test and our proposed modification (Eq. \ref{AD-Asym3}) with the aim to obtain some insight on the AD-Asym performances when the number of observations is small. We remark that for risk management purposes, as in the CCR case, the main goal is to reject more efficiently the forecasted (theoretical) distributions having a smaller variance than the actual (empirical) one (i.e. volatility underestimation).
We perform our numerical simulations assuming different theoretical probability density functions (PDFs). In particular, we compare a Gaussian distribution $N(0,1)$, assumed to be our theoretical PDF, with other Gaussian distributions having same mean but different variances. We estimate for a fixed number of observations the rejection rate of the AD test as a function of the real standard deviation considered to generate the sample set.
We expect that, for large enough sample size, AD test rejects the different distributions with a rejection rate equal to $1$ in $100\%$ of the cases, with the only exception when the real variance is equal to the theoretical one, i.e. equal to $1$. In the latter case, the rejection rate depends on the confidence level required for the test, in our case set equal to $5\%$.

\begin{figure}[h]
\centering
\subfloat[][\emph{AD-SS}.]
{\includegraphics[width=.45\textwidth]{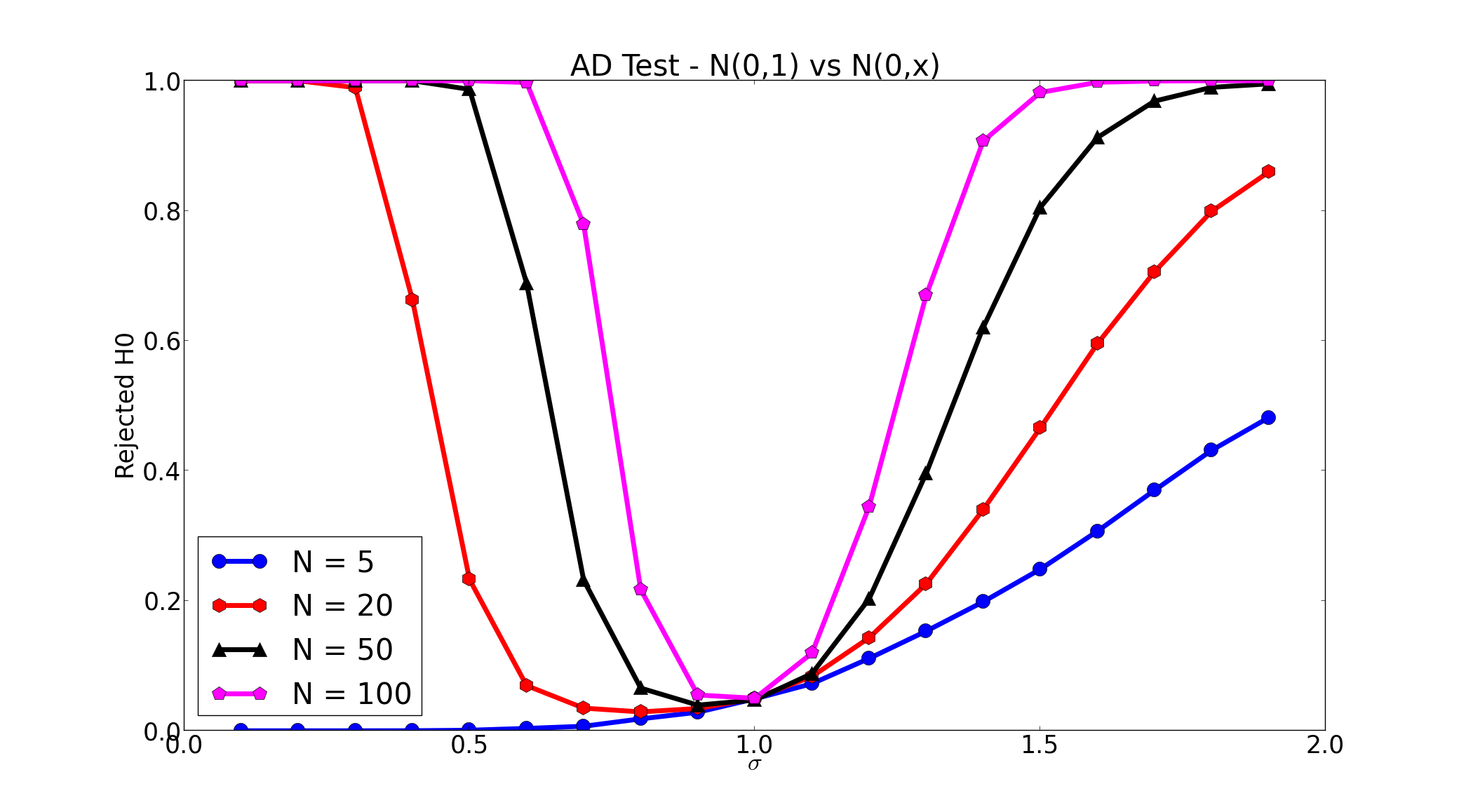}} \quad
\subfloat[][\emph{AD-Asym}.]
{\includegraphics[width=.45\textwidth]{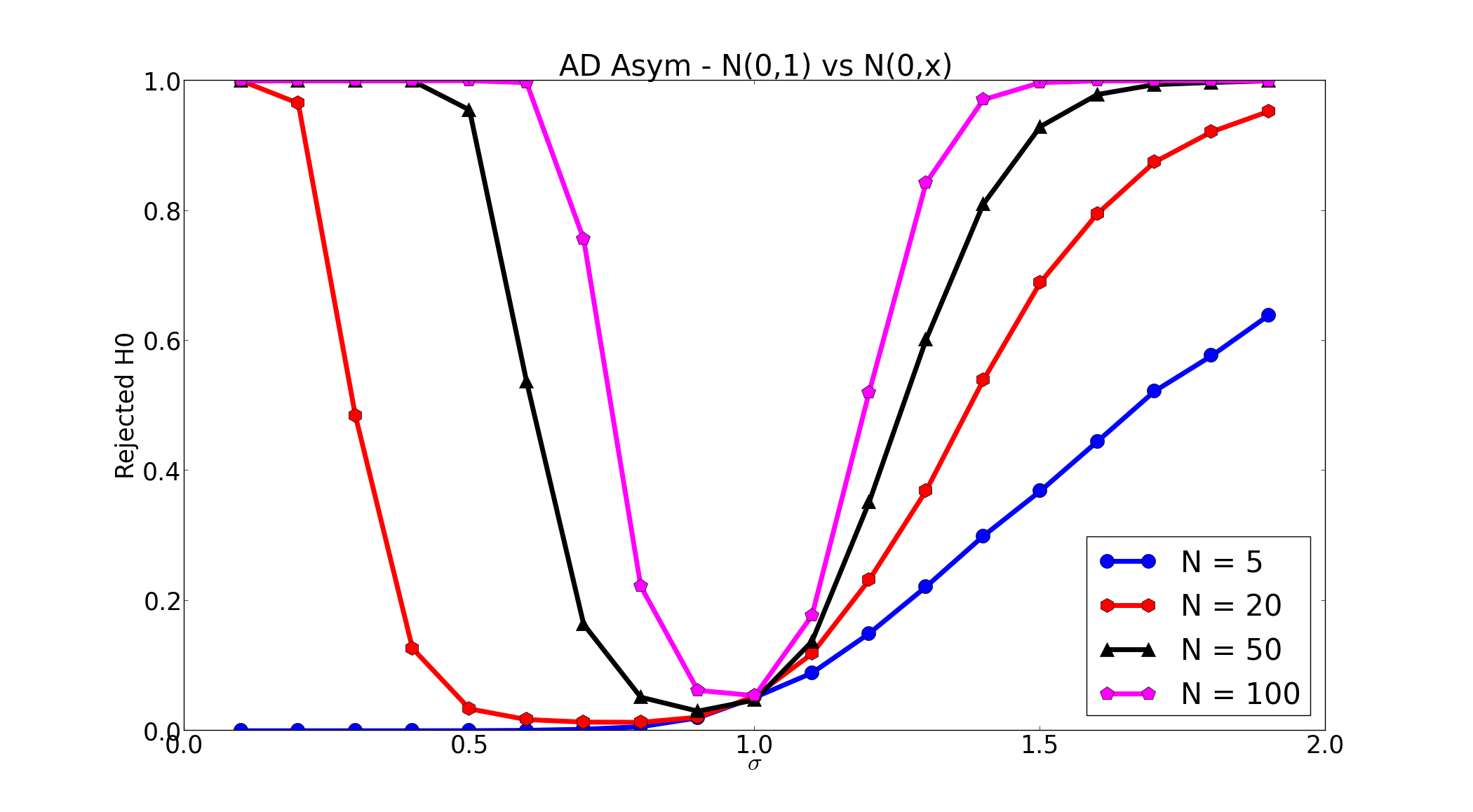}} \quad
\caption{Rejection rate as a function of the standard deviation}
\label{AD_AD-Asym}
\end{figure}

In Fig.~\ref{AD_AD-Asym} we report the results of our analysis for small sample size. As expected, the rejection rate is higher for larger standard deviations but the overall performance is not high. For example, we obtain a rejection rate around $50\%$ when considering $5$ observations and the standard deviation is twice larger than the empirical one. The asymmetry  between smaller and larger standard deviations becomes less and less evident as the number of observations increases: when $n=100$ the rejection rate becomes symmetric. 
 
\begin{figure}
\centering
\subfloat[][\emph{Sample size with 5 observations}.]
{\includegraphics[width=.45\textwidth]{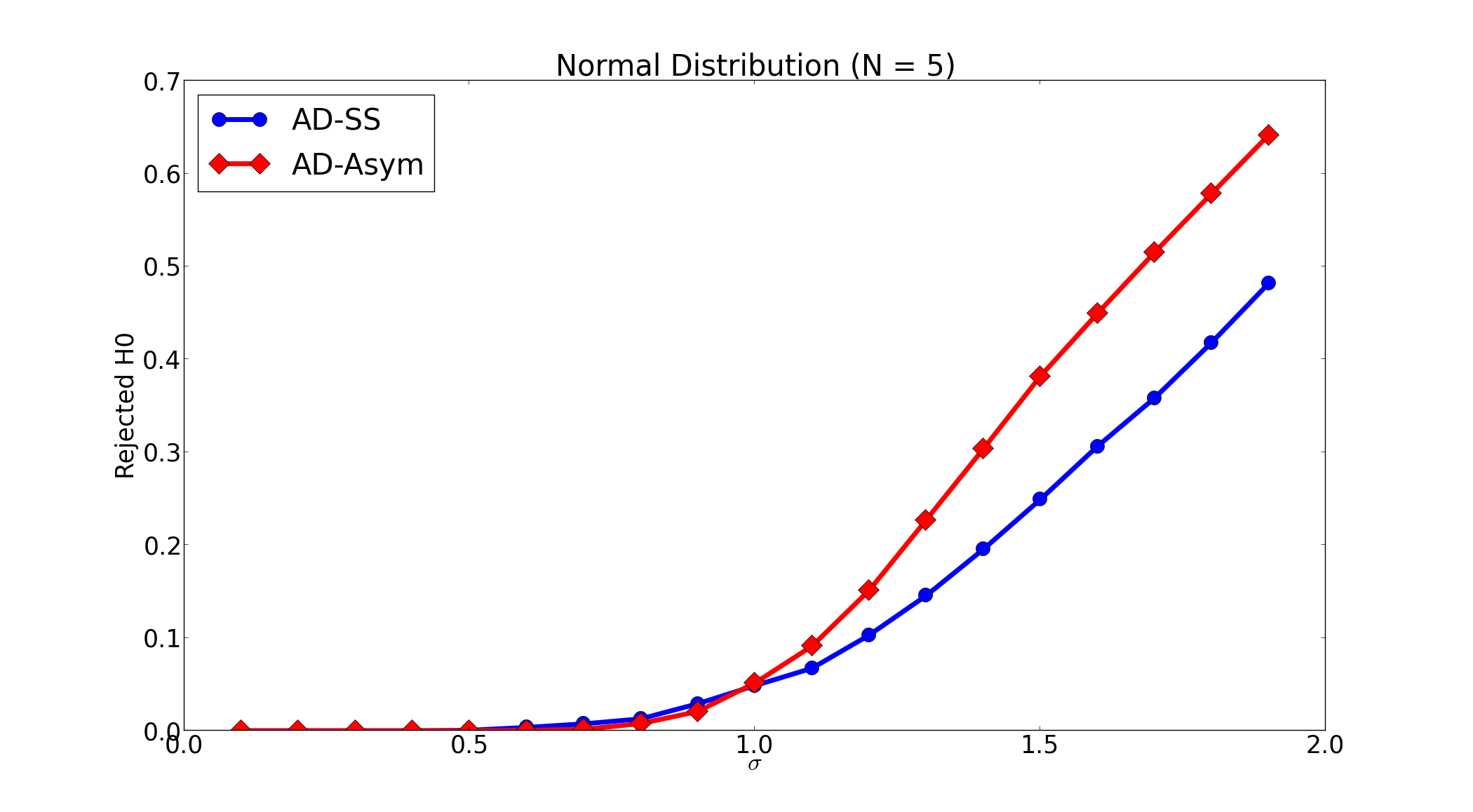}} \quad
\subfloat[][\emph{Sample size with 20 observations}.]
{\includegraphics[width=.45\textwidth]{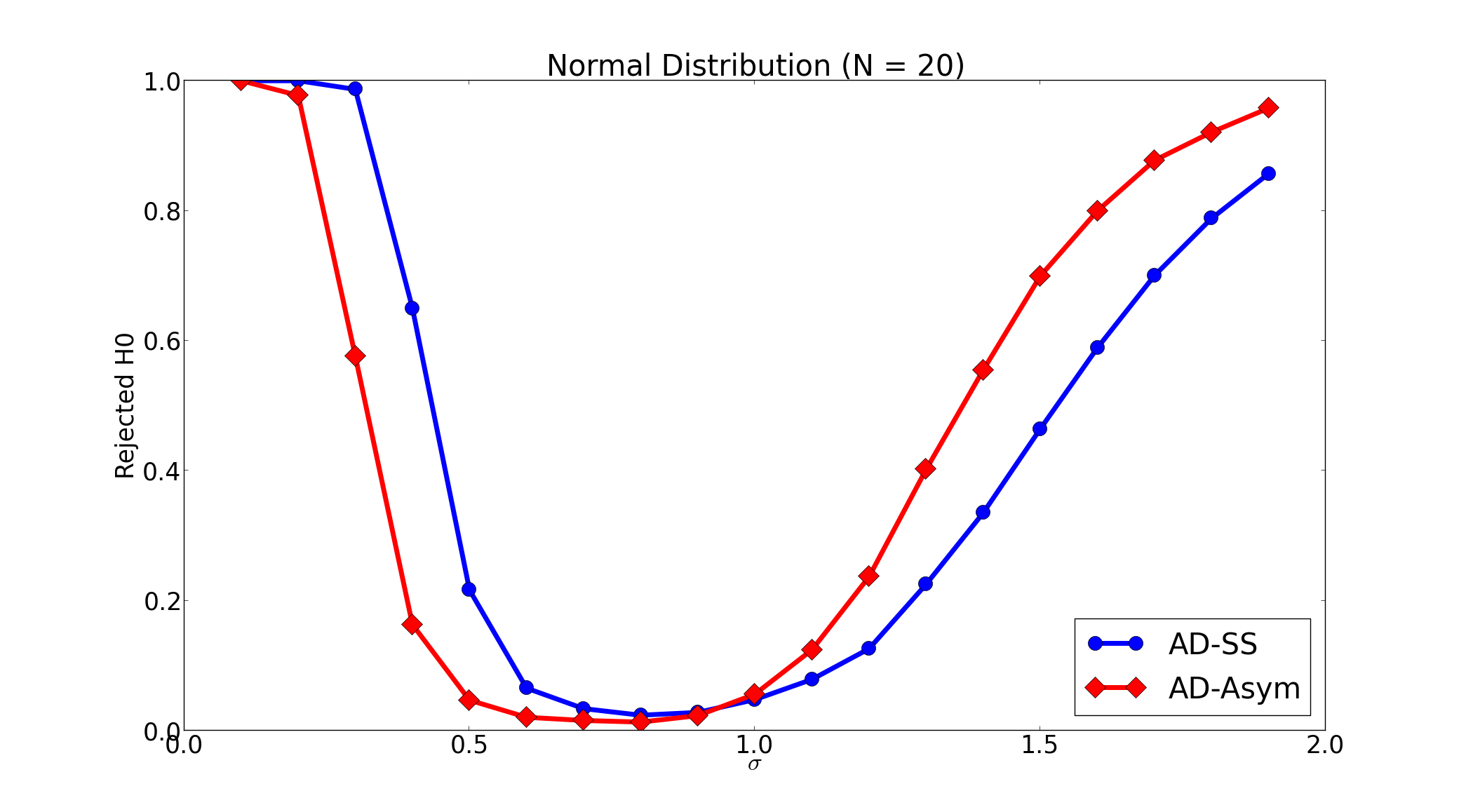}} \quad
\caption{Rejection rate as a function of the standard deviation}
\label{AD_AD-Asym-sample}
\end{figure}

In particular, the AD test performs better where we are more interested in (the right part of Fig.~\ref{AD_AD-Asym}-a), rejecting the most extreme cases when there is volatility underestimation. In fact, the AD test has a higher rejection rate when the distribution of real data has the same mean of the forecasted distribution, but a larger variance. The reason is, as already discussed in the previous section, that a realized PDF with a larger variance has a large probability to generate outliers (with respect to the forecasted PDF) even with few observations. On the contrary, if one considers the realized PDF with a lower variance than the forecasted one, all the observations will fall inside the forecasted distribution and it is more difficult to conclude if the two distributions are different. We replicate the same analysis comparing the AD test with our AD-Asym version using different sample size. Fig.~\ref{AD_AD-Asym}-b shows that we have a similar rejection rate when volatility is overestimated ($\sigma<1$) and a higher rejection rate when volatility is underestimated. Therefore, we conclude that if we need to avoid volatility underestimation the AD-Asym test performs better especially in limited sample. \\
Fig.~\ref{AD_AD-Asym-sample}-a plots the results when the sample size is made of five observations: the AD-Asym double rejects the null hypothesis when the volatility of the forecasted distribution is half the realized (e.g., AD-Asym rejects 40\% instead of 22\% when $\sigma$=1.5); the result is similar when observations are twenty as shown in Fig.~\ref{AD_AD-Asym-sample}-b.

In order to extend our result, We compare the rejection rate of different statistical tests aiming at identifying if a sample, that is always limited in our case, belongs to a given theoretical distribution. In particular, we considered five different statistical tests:
\begin{itemize}
\item Standard symmetric AD test, as described in the previous sections (SS)
\item AD test with tail sensitive indication, (AD-Asym)
\item Cramer-Von Mises (CM)
\item Kolmogorov-Smirnov (KS)
\end{itemize} 

\begin{figure}
\centering
\subfloat[][\emph{Tests assuming volatility overestimation}.]
{\includegraphics[width=.45\textwidth]{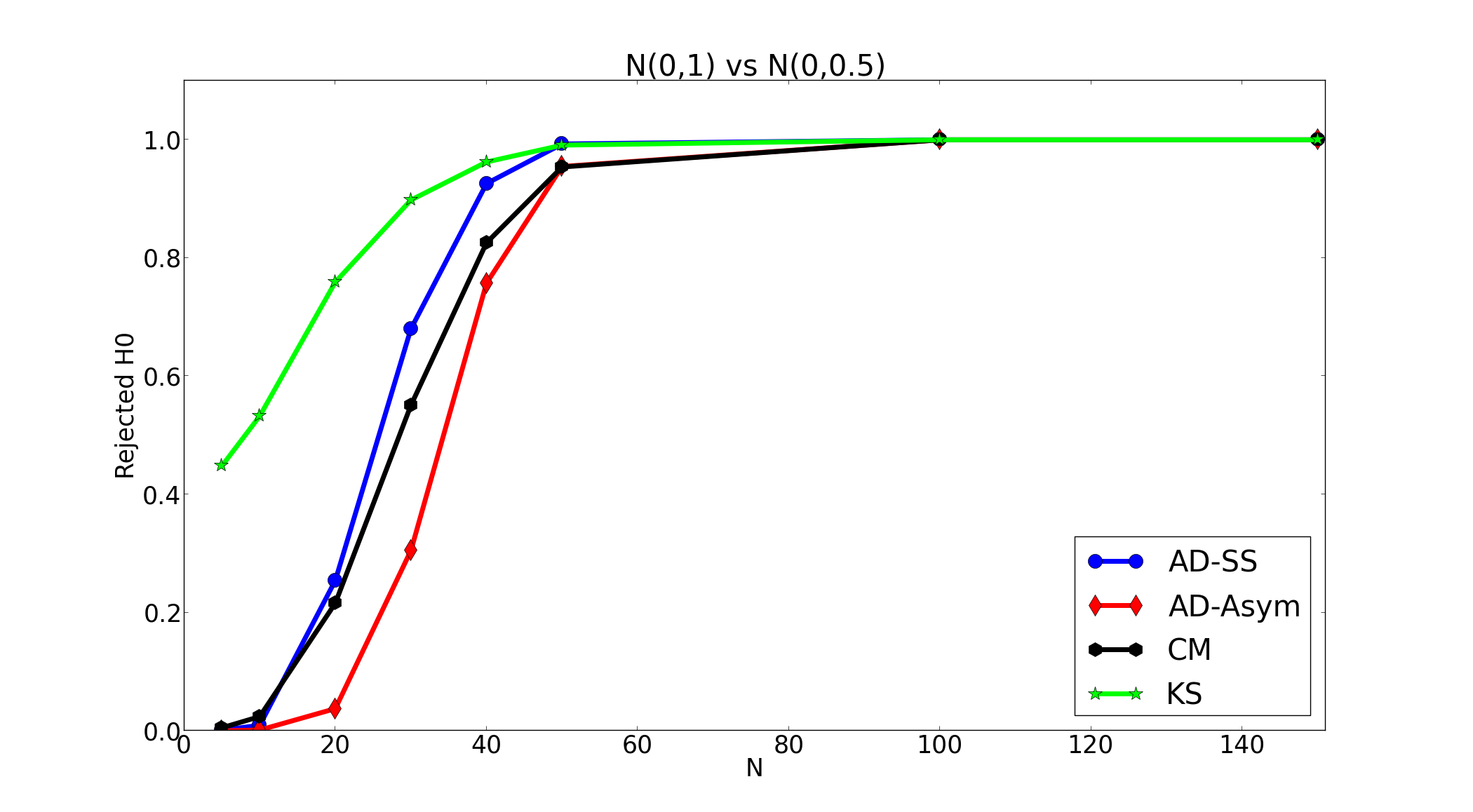}} \quad
\subfloat[][\emph{Tests assuming volatility underestimation}.]
{\includegraphics[width=.45\textwidth]{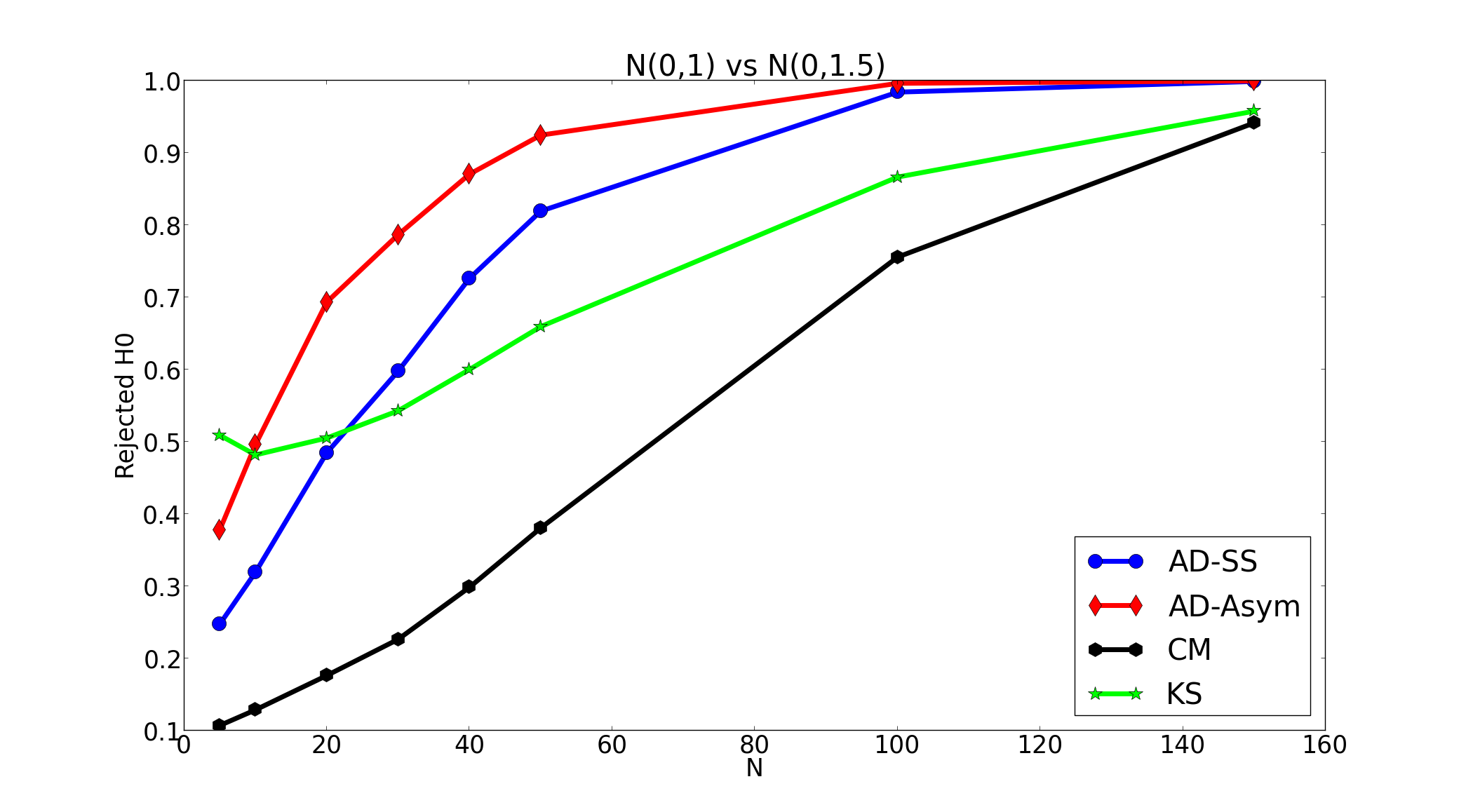}} \quad
\subfloat[][\emph{Tests assuming mean overestimation and the correct volatility}.]
{\includegraphics[width=.45\textwidth]{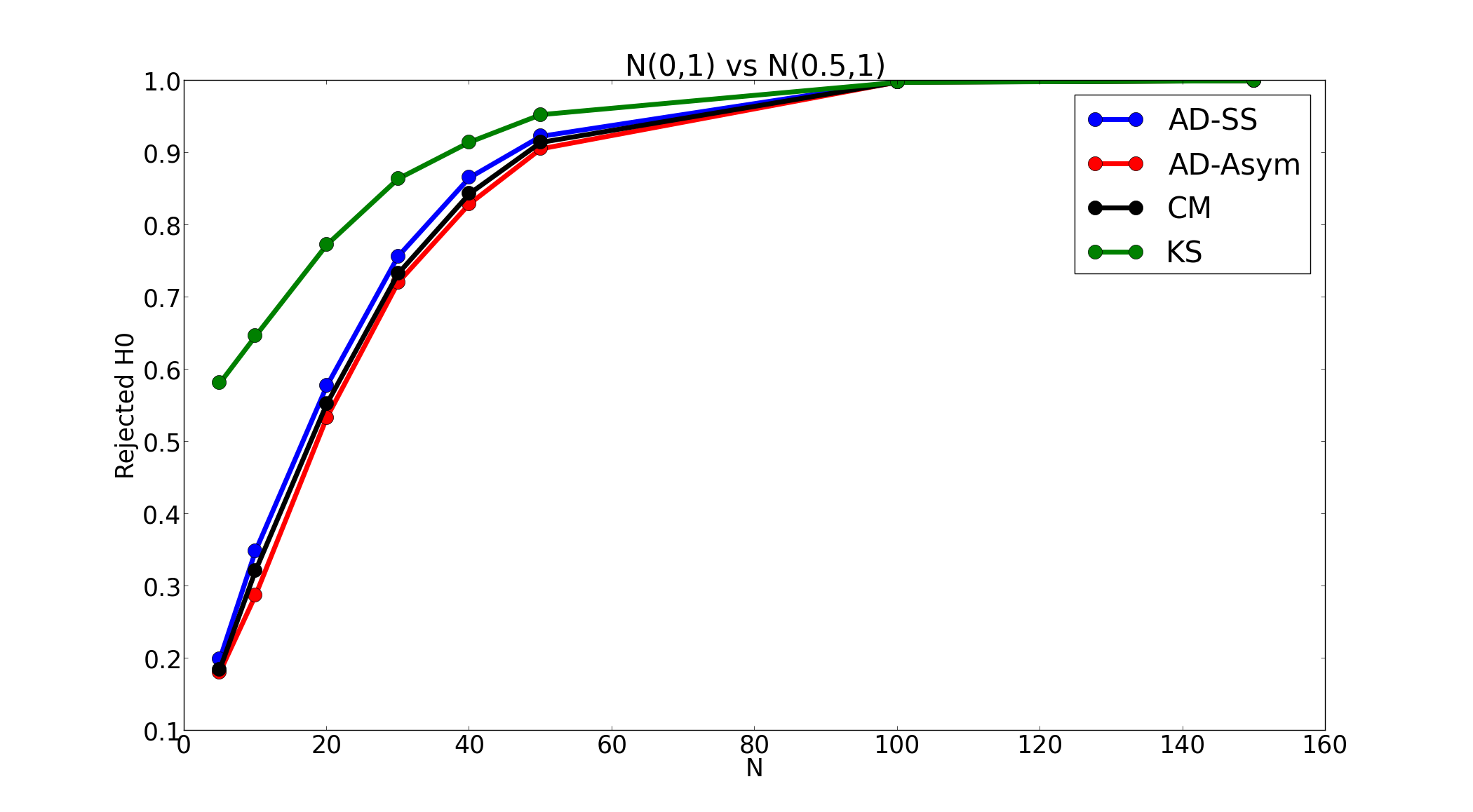}} \quad
\subfloat[][\emph{Tests assuming mean overestimation and volatility underestimation}.]
{\includegraphics[width=.45\textwidth]{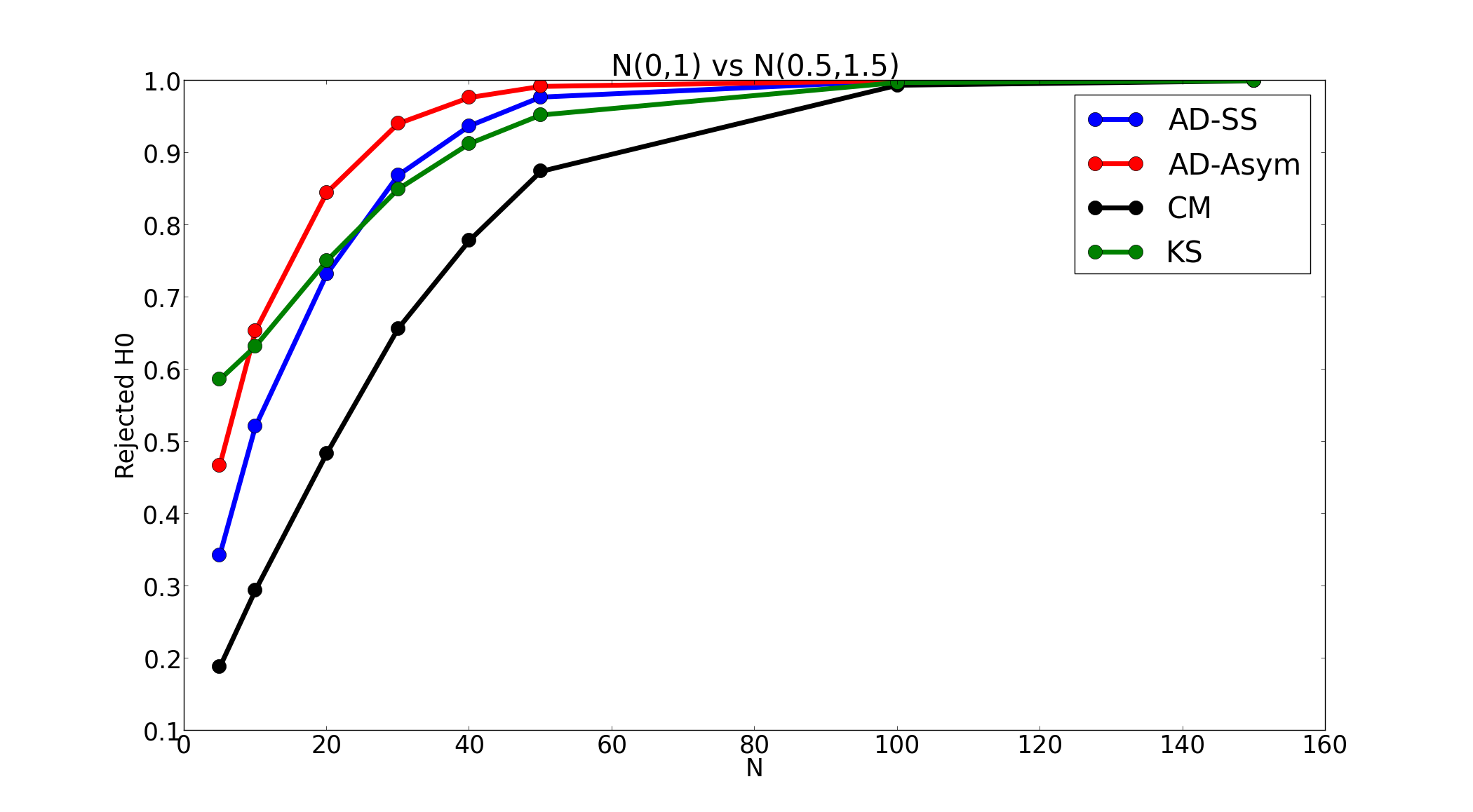}}
\caption{Rejection rate of different statistical tests}
\label{Analysis}
\end{figure}

Fig.~\ref{Analysis} shows the results of our analyses as a function of $n$; in particular, in each figure, the rejection rate of different statistical tests is compared assuming a Gaussian $N(0,1)$ theoretical distribution and, empirical Gaussian distributions with different means and standard deviations. As expected, AD-Asym test shows the overall better performances (higher rejection rate) than the other tests, when the volatility is underestimated.

\begin{figure}[h]
\centering
\subfloat[][\emph{t-Student with $\nu=2.8$ degrees of freedom}.]
{\includegraphics[width=.45\textwidth]{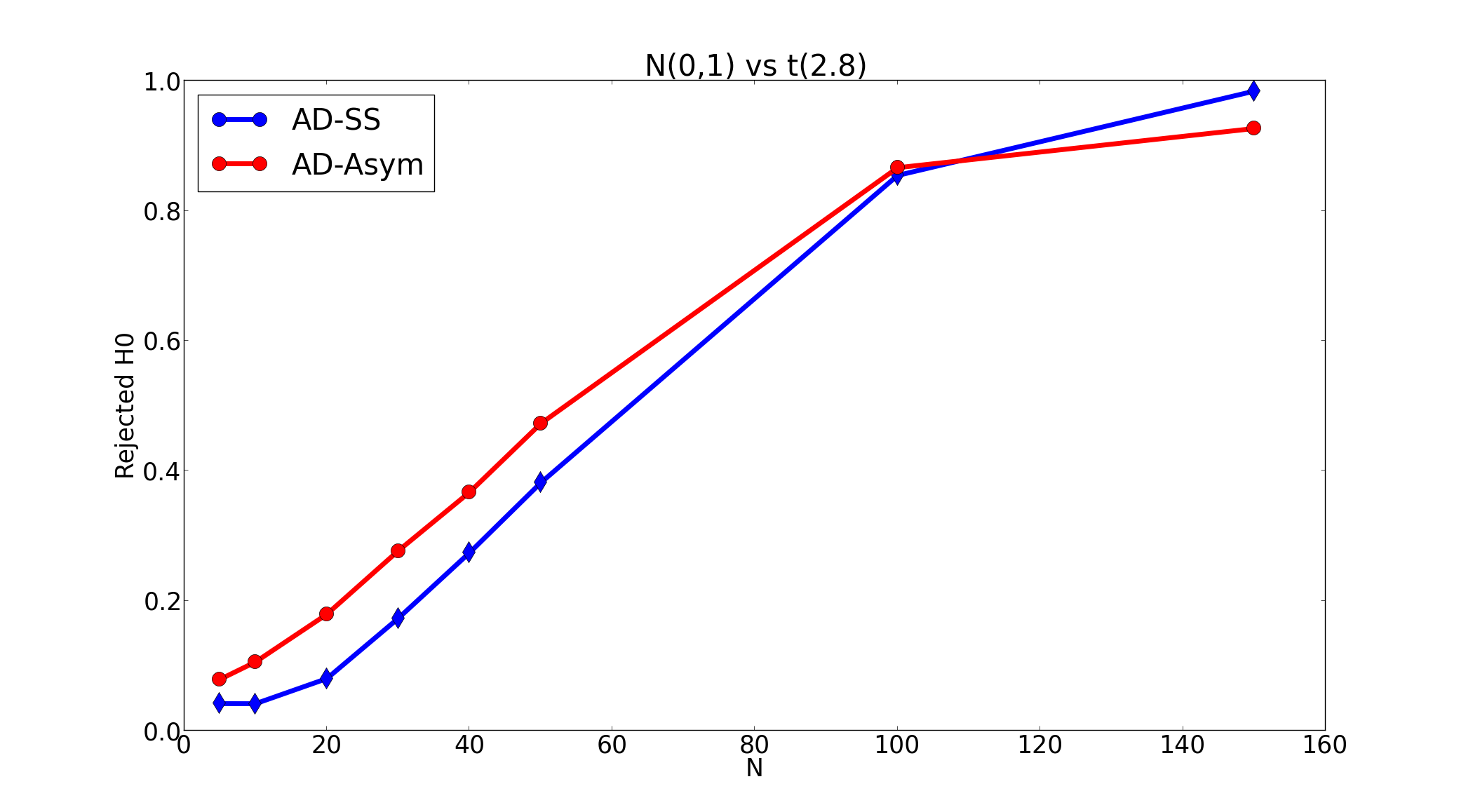}} \quad
\subfloat[][\emph{t-Student with $\nu=3$ degrees of freedom}.]
{\includegraphics[width=.45\textwidth]{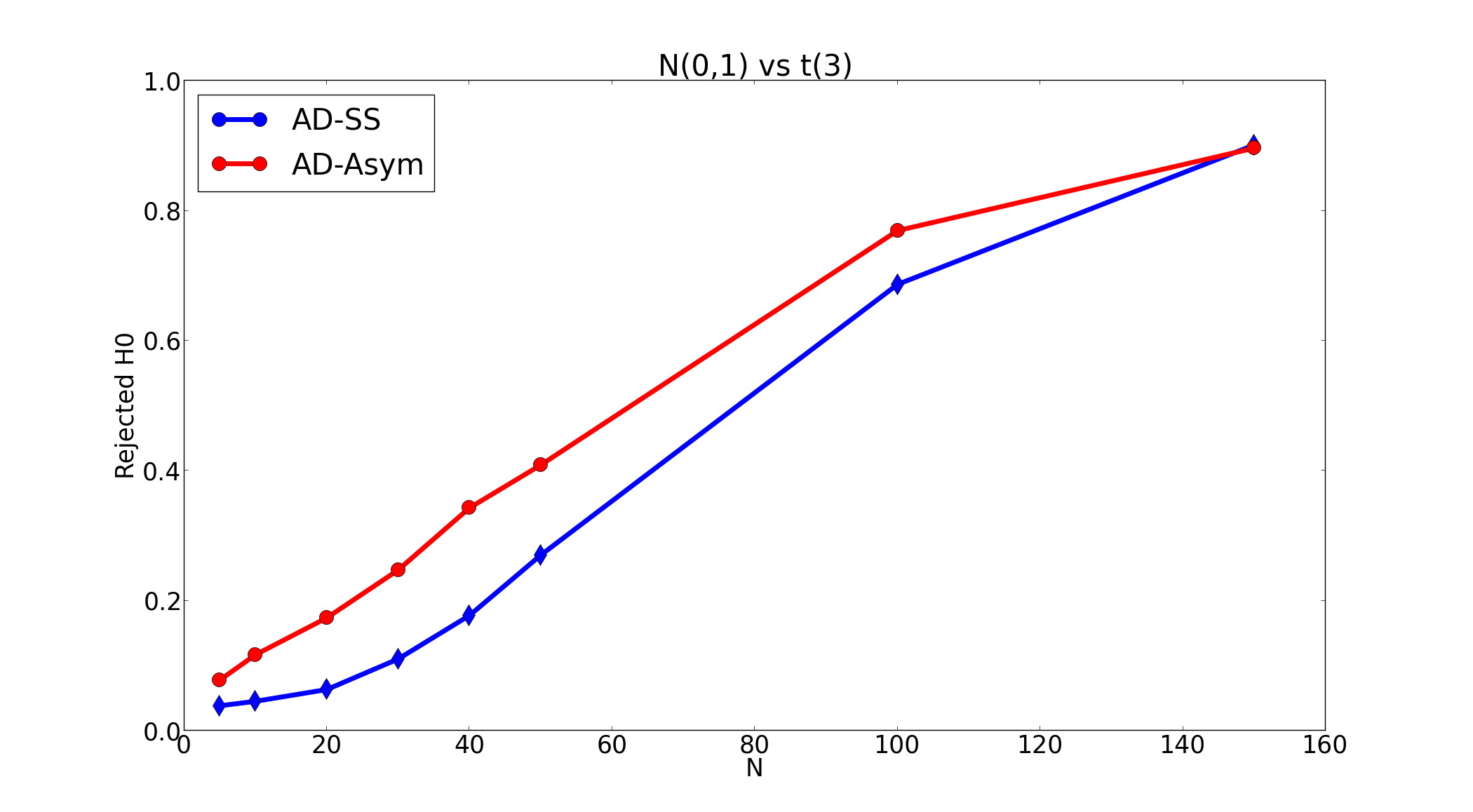}} \quad
\subfloat[][\emph{t-Student with $\nu=3.5$ degrees of freedom}.]
{\includegraphics[width=.45\textwidth]{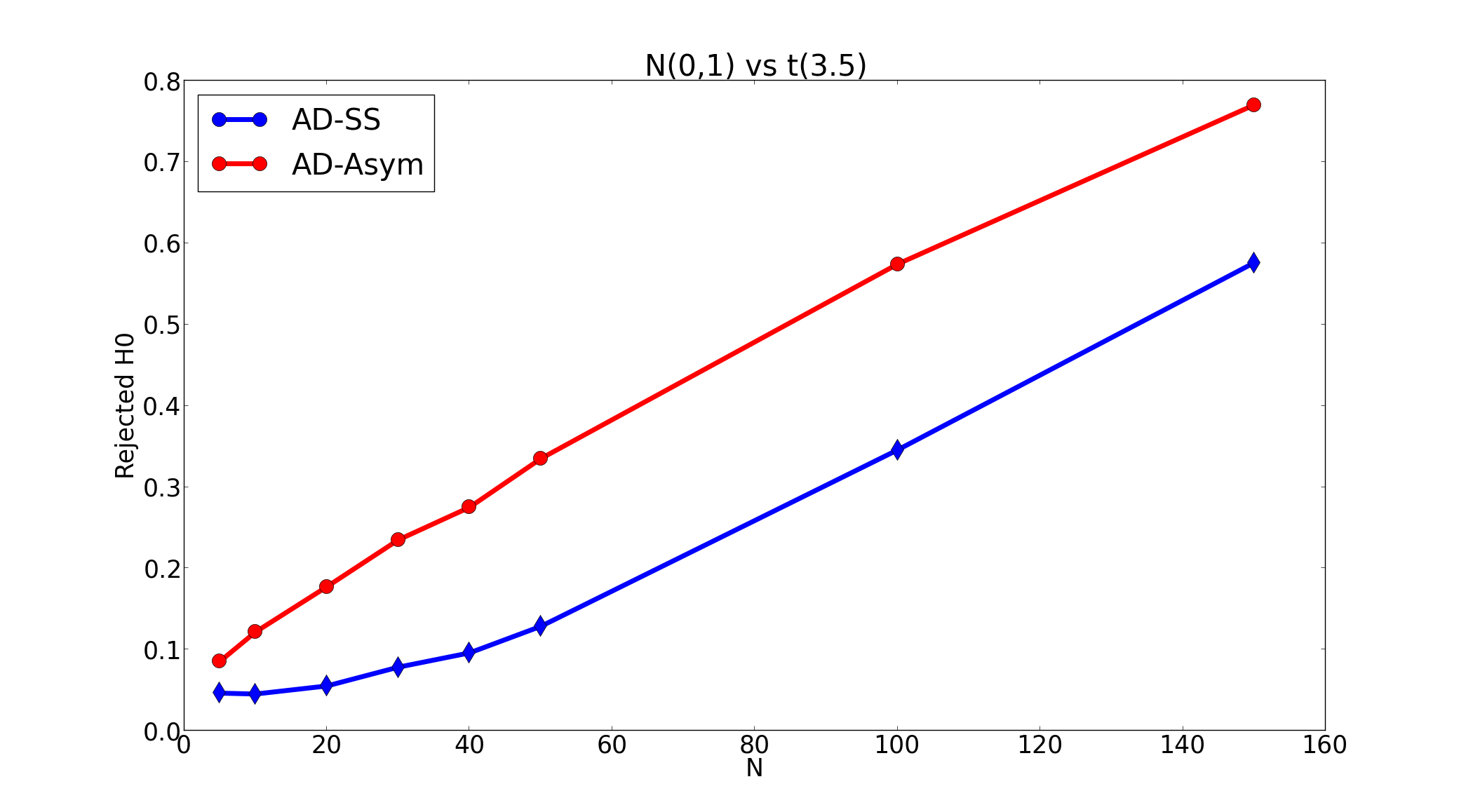}} \quad
\caption{Rejection rate for AD test (blue line with diamonds) and AD-Asym test (red line win dots) assuming a Gaussian N(0,1) theoretical distribution and a t-Student distribution}
\label{Comparison}
\end{figure}

As it is a well-known fact that the distributions of financial returns frequently show fat tails behavior, we analysed the performances of different statistical tests considering a Gaussian N(0,1) theoretical distribution and an empirical t-Student distribution with variance equal to one and $\nu$ degrees of freedom. In Figs.~\ref{Comparison} we show the results of our analyses.
The overall result is that AD-Asym performances are higher than the ones related to the AD test, although we know such test are strongly influenced by the degrees of freedom considered. This fact can be explained considering the fat-tails nature of the t-Student distribution that implies an higher risk if compared to a Gaussian distribution with the same variance. For this reason, the use of a Gaussian distribution would imply an underestimation of the risk that is better identified by AD-Asym test, as already spotted out in the previous section. In addition, we observe that the convergence of the rejection rate to $100\%$ is not reached when $n=150$, on the contrary to the Gaussian case. The reason for this behaviour can be found if one considers that, as the first two moments (mean and the variance) of the theoretical and the empirical distribution are exactly met, one need to estimate the higher moments in order to observe differences between the PDFs. 
Given that for a fixed number of observations the uncertainty increases with the moment order, it is reasonable to expect that the number of observations needed to obtain the convergence must be larger than the one obtained with Gaussian PDFs (see section~\ref{analytical}).

\section{Numerical Exercise}
\label{NumResults}
\begin{figure}[t]
\centering
\subfloat[][\emph{Time Horizon: 1 Week}.]
{\includegraphics[width=.4\textwidth]{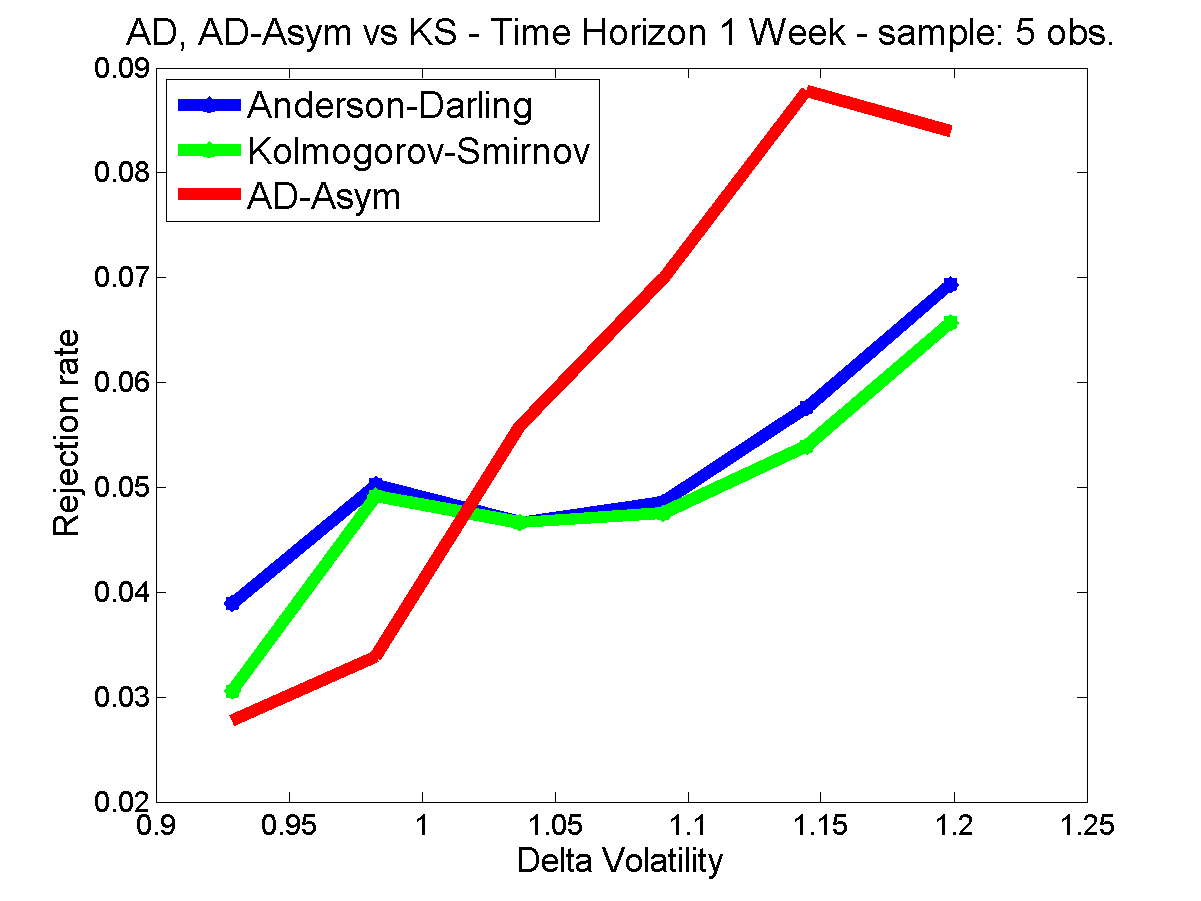}} \quad
\subfloat[][\emph{Time Horizon: 1 Month}.]
{\includegraphics[width=.4\textwidth]{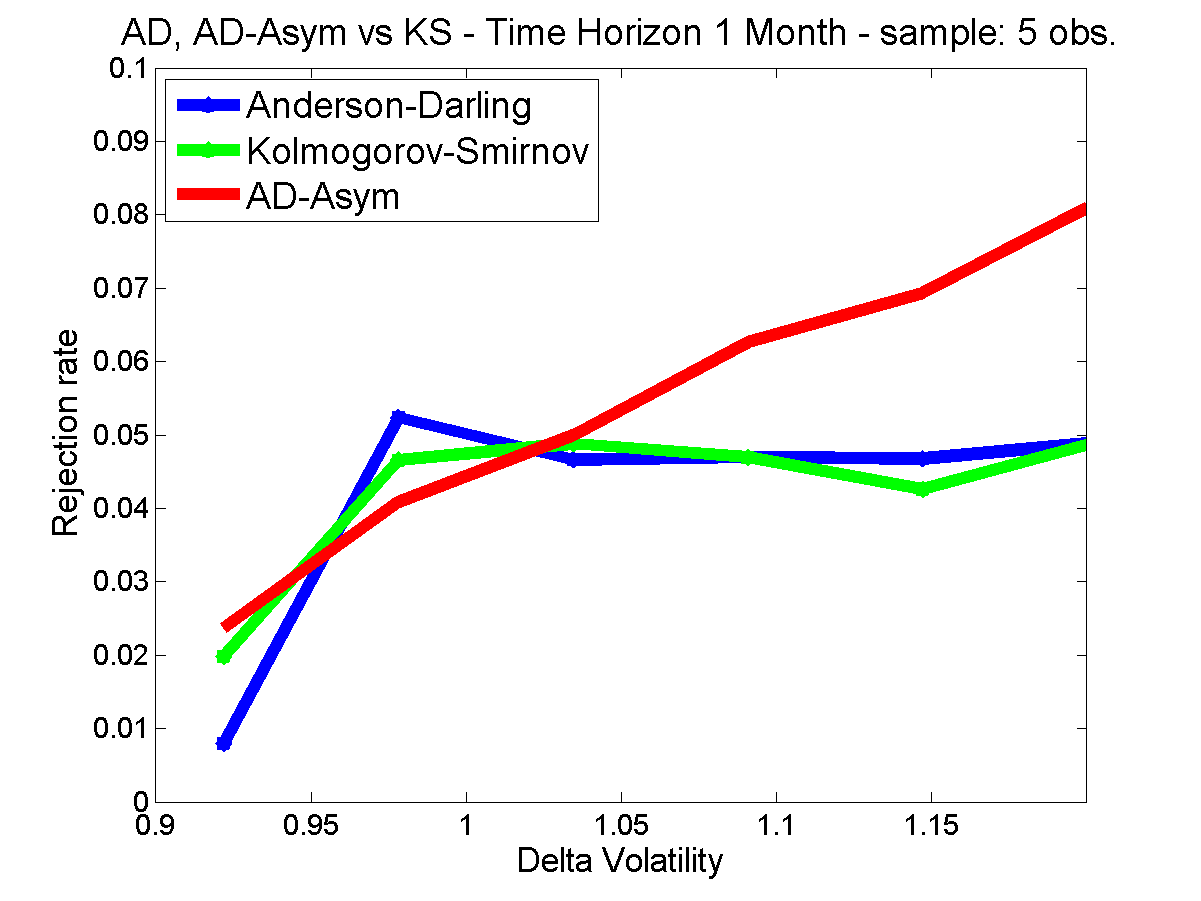}} \quad
\bigskip
\subfloat[][\emph{Time Horizon: 1 Year}.]
{\includegraphics[width=.4\textwidth]{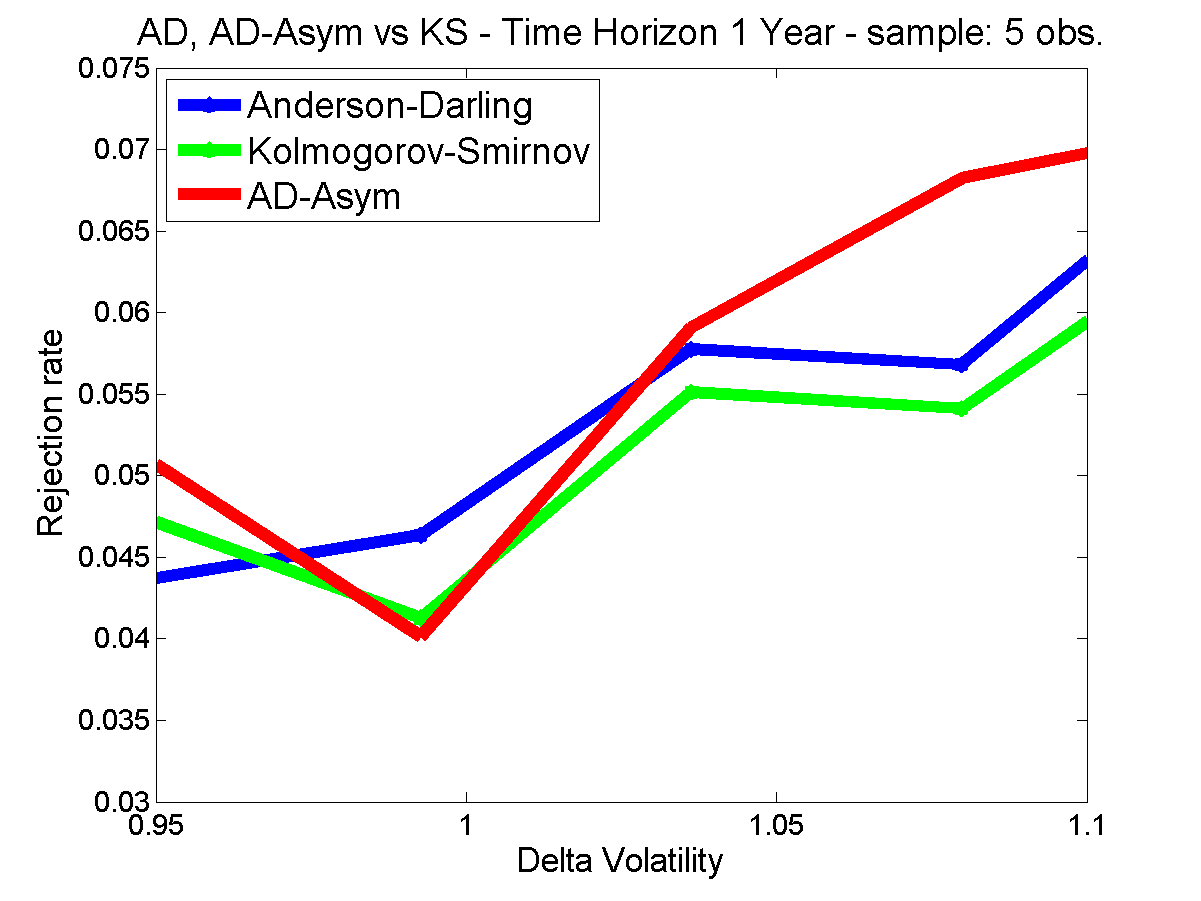}} \quad
\subfloat[][\emph{Time Horizon: 2 Years}.]
{\includegraphics[width=.4\textwidth]{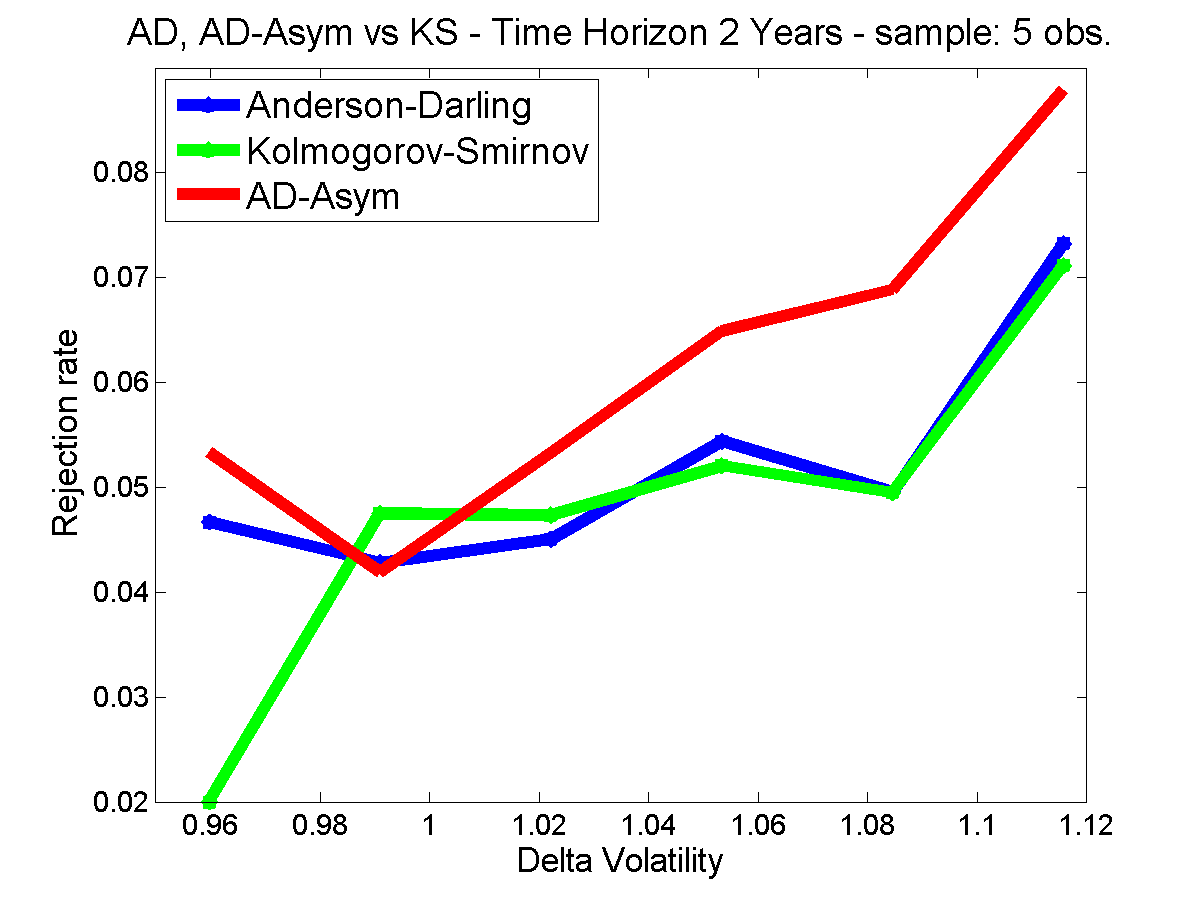}}
\caption{AD vs KS rejection rate using 5 observations in the backtesting sample}
\label{pippo}
\end{figure}

Once understood the main effect from a theoretical point of view, we test the AD in a realistic case. We consider the Euribor 6-month interest rate having historical time series since 1999 and, as a consequence, it has five/six observations in the backtesting sample, if we are using three years of historical data for estimation purposes and a forecasting horizon of two years. 
However, in order to address the property of AD test in limited sample size but having the possibility to extend our analysis to higher sample size, we detrend the interest rate time series using the Hodrick-Prescott~\cite{Hodrick} filter applied to the five days time series up to March 2012. 
In this way, we were able to obtain more robust calibration results of interest rate model.
We then estimate a Black-Karasinski~\cite{bk} interest rate model on the cycle component of the log-filtered time series in order to generate, using that estimated parameters, 'fictitious' time series with the desired sample to backtest. 
We use the Black-Karasinski  short-rate model because of its analytical tractability and for the added benefit that rates cannot become negative. In fact, the Black-Karasinski models the logarithm of the cyclical component of the interest rate $y(t)=\mathrm{exp}(r(t))$ using the Ornstein-Uhlenbeck stochastic process:
\begin{equation}
\label{BK_eq}
\mathrm{d}y_{t} = k \bigl( \alpha - y_{t}\bigr) \mathrm{d}t + \sigma \mathrm{d}W_{t}
\end{equation}
where $k$ captures the speed of the $y_{t}$ log-Cycle toward its long equilibrium value $\alpha$ (the mean level), $\sigma$ is the volatility of the process and $W_{t}$ is the brownian motion. We estimate the following parameters using a moment matching approximation formula method on the overall dataset: 
\[
\Theta \equiv [\tilde{\alpha}=-0.0004871; \quad\quad \tilde{k}=0.011853; \quad\quad \tilde{\sigma}=0.018855]\\
\]
The moment matching method guarantees the level of mean reversion is positive, which is consistent with the values of the risk factor. In particular, the estimation method is based on a two-steps formulas where the mean reversion level is firstly estimated and then plugged in the mean reversion rate:  
\begin{eqnarray}
\label{ergodic}
\alpha 	&= & \frac{1}{n}\sum_{k=1}^{n}y(t_{k}) \\
k		&= & \frac{\sum_{k=1}^{n}\bigl(y(t_{k})-y(t_{k-1})\bigr)^{2}}{2 \sum_{k=1}^{n}\bigl(y(t_{k}) - a\bigr)^{2}}
\end{eqnarray}
\begin{figure}[h]
\centering
\includegraphics[width=.95\textwidth]{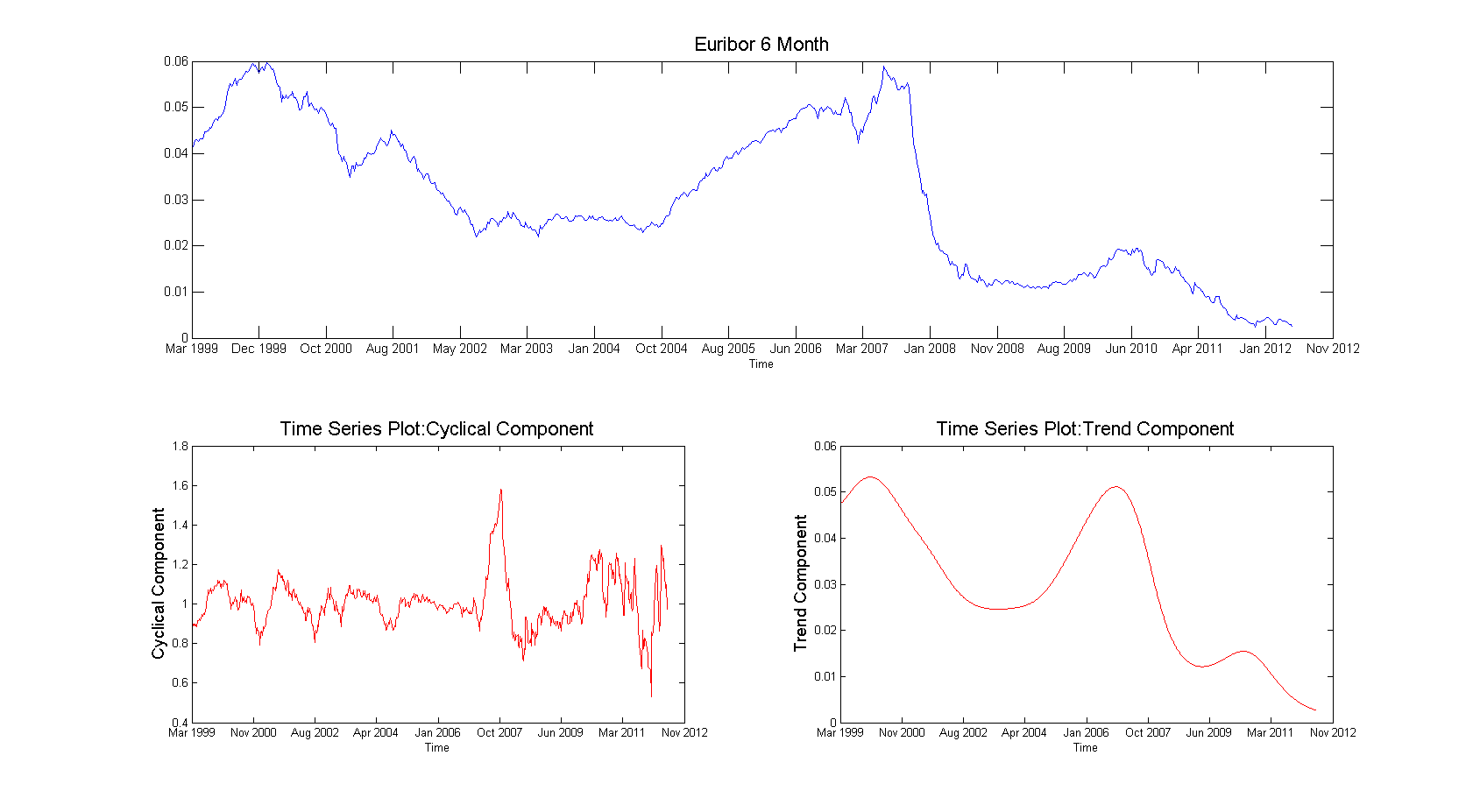}
\caption{Interest rate Euribor 6 months filtered out with Hodrick-Prescott and simulations using the estimated parameters $[\tilde{\alpha}, \tilde{k}, \tilde{\sigma}]$}
\label{Eur-6m-filter}
\end{figure}
In Fig.\ref{Eur-6m-filter} we plot the Euribor 6 Months, the filtered cyclical and the trend component.
We simulate interest rates scenarios using the estimated $\Theta$ parameters and a constant number of backtested observations, in order to verify the statistical properties of the tests at different time horizons. 
In the Fig.~\ref{pippo} we show the results of the AD and KS rejection rate using at every simulation a backtesting sample with 5 observations. All simulations are performed 10 thousand times using the same parameters. The rejection rate is plotted against the ratio between the average standard deviation of the simulated process (using $\Theta$ parameter) and the standard deviation at every backtesting dates (calibrated considering a time window of three-years and keeping $k$ constant\footnote{The variance of the process, as described in Appendix B, depends on the mean reversion rate $k$ so in order to avoid a misalignments between the volatility used to simulate the process ($\Theta$) and the ones used for backtesting purposes, we opt to take a constant $k$, with an overall effect that we verified is negligible for our purposes.}):
\[
\Delta=\tilde{\sigma}_{sim}/\tilde{\sigma}_{bkt}  
\]
\begin{figure}
\centering
\includegraphics[width=.7\textwidth]{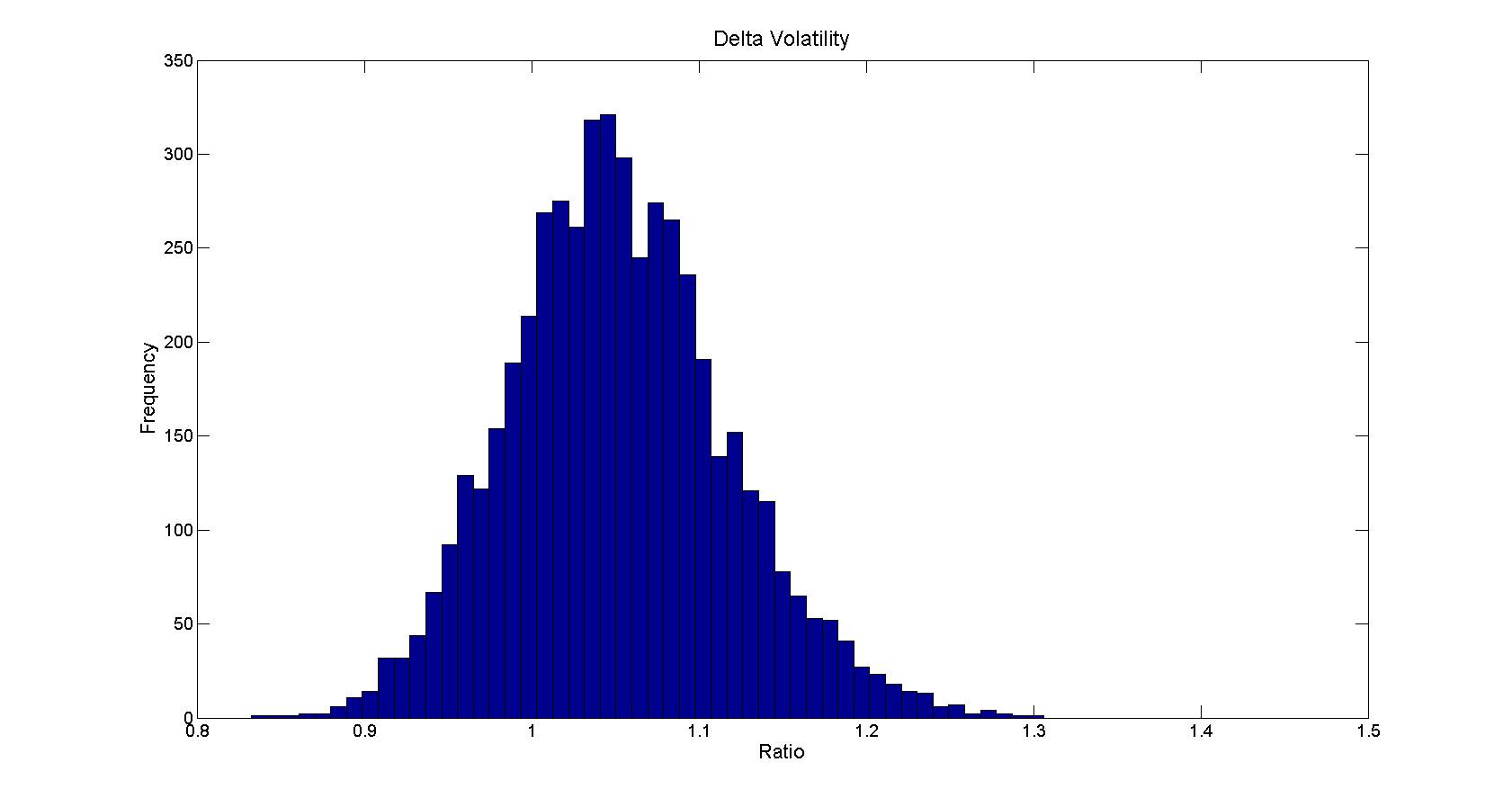}
\caption{Distribution of $\Delta=\tilde{\sigma_{sim}}/\tilde{\sigma_{bkt}}$  }
\label{volatility_histogram}
\end{figure}
where the variance of the process is estimated as described in Appendix \ref{appB}.\footnote{The variance of the Black-Karasinski process depends on the time horizon, so we expect such difference on the x-axis as shown in Figures \ref{pippo}} As a consequence, a positive $\Delta$ implies an underestimation of the volatility of the backtesting sub-samples, therefore we expect a higher rejection rate of both tests, while a negative $\Delta$ implies the backtesting volatility is higher than the one that generated the 'fictitious' time series, so we expect a lower or even a zero rejection rate. The numerical exercise seems to confirm the theoretical analysis shown in Figs. \ref{fig_coefficient_variation}-\ref{AD_AD-Asym} where the AD test has a higher rejection rate compared to KS when sample size is low.\footnote{The extreme boundaries of Fig. \ref{pippo} have few observations so we discard it from our analysis.} We control these results using 10, 25 and 100 observations obtaining similar results. Last, we notice that the volatility estimated using three years is on average lower than the one used for simulating the 'fictitious' real interest rate. This is confirmed in Fig. \ref{volatility_histogram} where we plot the histogram of the distribution of $\Delta$.

\section{An empirical application of Anderson-Darling test}
\label{Empirical Results}
In this section, we show an example of a risk factor backtesting applying the AD and KS test on interest rate forecasted values obtained using the Black-Karasinski \cite{bk} short-rate model. We computed this exercise to show a real test case of forecasting long time horizon and, at the same time, to verify the performance of AD test in real case.

\begin{figure}[h]
\centering
\includegraphics[scale= 0.5]{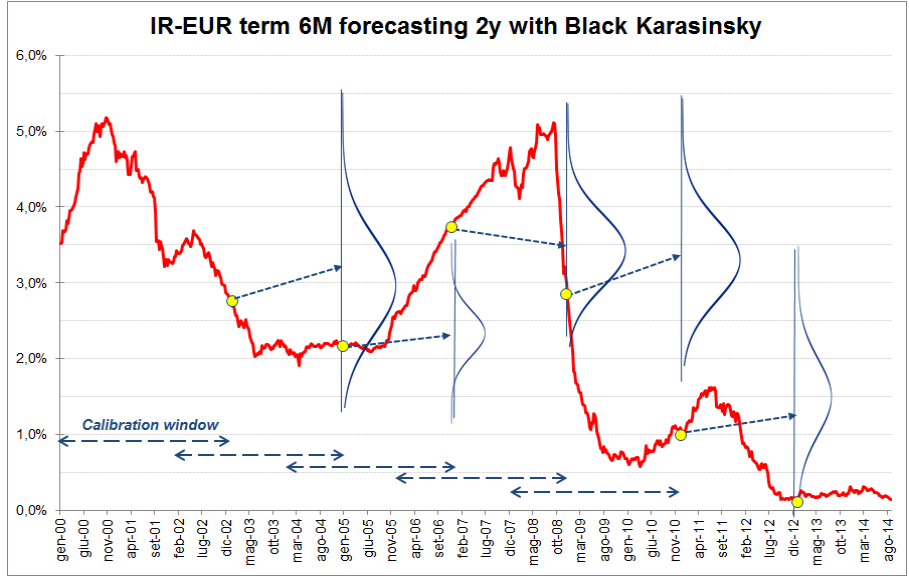}
\caption{Euribor 6-Months: forecasting distribution at 2 years time horizon}
\label{IREUR_6M_Forward}
\end{figure}

We still remark that  we are interested in backtesting mark to future distribution at different horizon since regulatory measurements
and managerial exposures are derived from the entire distribution of interest rates products. For simplicity reason, we still use the Black-Karasinski as given by equation (\ref{BK_eq}) applied to the Euribor 6-Month time series where parameters are calibrated using 3 years of historical data and the moment matching method shown in equation (\ref{ergodic}). 
We slightly modify the short rate model (\ref{BK_eq}), in order to perform tests with alternative model set-up, as follow
\begin{equation}
\label{BK_eq_adj}
d \, y_{t} = k \bigl( \alpha(t) - y_{t}\bigr) dt + \gamma \sigma dW_{t}
\end{equation}
where the parameter $\gamma$ is an adjustment to the volatility value. Fig. \ref{IREUR_6M_Forward} plots the Euribor 6-Months and the forecasted distribution at two years when setting $\gamma=1$\footnote{We use 3000 simulations at each backtesting dates.}. We remark that the backtesting time window covered the interest rate regime switch occurred during the Lehman crisis, and in general interest rate models would have serious problems to correctly forecast the crisis and the following period. 
Fig. \ref{IREUR_6M_Forward} shows that: (i) only five backtesting dates are available for testing the model due to the long forecast horizons; (ii) the empirical data often fall in the extreme tail of the forecasted distribution; (iii) only by increasing model volatility or adjusting the mean one could capture the realised values.
Table~\ref{TabFormenti1} reports the simulated rates and the corresponding real rates at the five backtesting dates.
\begin{table}[h]
\centering
\begin{tabular}{l c c c c c }
	\hline\hline
	Date 		&  24 dec '04 & 22 dec '06 	& 19 dec 2008 	& 17 dec '10 & 14 dec '12 \\
	\hline
	Value		& 2.211		  & 3.8291    	& 2.7634 		&	1.0449 	 & 0.1685  \\ 
	\hline
	Minimum		& 2.0609      & 1.2958      & 2.2502     	&  1.8545    & 0.16382  \\
	Average		& 3.3743      & 2.3433      & 3.4989     	& 3.3177     & 1.2567  \\
	Maximum		& 5.5148      & 3.7534      & 4.9848      	& 5.357      & 4.8442 \\
	\hline 
	\hline
\end{tabular}
\caption{Euribor 6-Month and statistics forecasted values at two years time horizon}
\label{TabFormenti1}
\end{table} 
Table~\ref{TabFormenti2} reports the results of the AD, AD-Asym and KS tests for different level of volatility assuming a 5\% level of confidence\footnote{A test rejects the null hypothesis when the p-value is lower than the confidence level.}: all tests reject the hypothesis that the model is correct (see the bold row and column). In order to compare the performance of AD, AD-Asym and KS test we re-parametrized the model by increasing volatility. In particular,  we increase $\gamma$ from one up to three times the estimated volatility at each backtesting date and we observe that, as expected, tests accept the null hypothesis when the forecasted distribution is very large. On the contrary the AD-Asym is less reactive to such adjustments compared to AD and KS. We consider it as an empirical proof of conservativeness of our test. 
Last, we check the robustness of those results with different term of the curve (i.e, 1 year, 5 year and 10 year term) obtaining similar results. 
\begin{table}[h]
	\centering
	\begin{tabular}{l c c c c c c c c }
		\hline
		&Test/$\gamma$ 	& 1  		  		& 2      	& 2.5  			&	2.75 	& 3 \\
		\hline\hline 
		& AD 			& \textbf{Reject} & Reject  	& Reject 		& Accept 	& Accept \\
		& AD-Asym		& \textbf{Reject} & Reject  	& Reject		& Reject 	& Accept \\
		& KS  			& \textbf{Reject} & Reject	 	& Reject		& Accept 	& Accept \\
	\hline
	\hline
\end{tabular}
\caption{AD, AD-Asym and KS results for Euribor 6-Month forecasted values at 2 year time horizon.}
\label{TabFormenti2}
\end{table}


\section{Conclusion}
\label{conclusion}
The European CCR/CRD IV requires to perform an in-depth analysis of model's forecasts used to compute the Counterparty Credit Risk exposures. As a consequence, banks need to perform a backtesting program which results are very important for the assessment of model weaknesses that impact on counterparty exposures and risk weighted assets. On the same time, banks face a common situation when forecasting risk factors at long time horizons but satisfying the requirement of three years data for model calibration: a very limited sample dataset. With this respect, from a purely statistical point of view, the use of overlapping time widows to verify model performance cannot be considered a significant improvement for what concern the reduction of statistical uncertainty of the forecasted variables. As a consequence, a proper approach for the CCR model backtesting should be developed taking into consideration the strong constrains given by the limited statistical power of small dataset. With this respect it worths to observe that, in general, the statistical test should be able to detect easily the model's volatility underestimation, that is more dangerous from a risk management point of view, than volatility overestimation.

For these reasons, in this work we focused on the Anderson-Darling (AD) test with the goal to understand when test is able to detect a volatility underestimation, and we derive an extension of the AD test that can be used for risk management purposes. The main idea for such extension relies on the observation that AD test rejects more the case where the real variance of the distribution is larger than the forecasted one, but the rejection rate is in any way low due to the uncertainty related to the small sample size. As a consequence, we magnify this asymmetric effect introducing a more pronounced non linear behavior when the differences between the forecasted and the real distribution are large when the model's volatility is underestimated. This means that test rejection power will be higher when the
forecasted distributions has a smaller variance than the real one. And this property should also hold in case of limited sample size in
order to be conservative.

We verified the property of our modified test in the limited sample size case showing that it has an overall better rejection performance in comparison to the standard uniform test when the forecasting distribution is wrong. We check this result comparing it with the AD test and other uniform test, such as the Kolmogorov-Smirnnov test, using a numerical example that forecasts the interest rate Euribor 6-Month values at two year time horizon with a Black-Karasinski short rate model. We then use such model to backtest the real interest rate during the time window 2000-2014 that lead us to reject the chosen model.

\newpage{}
\appendix
\section{Functions Definition for AD-Asymmetric Test}
In the following, we report the functions defined in Eq.~\eqref{AD-Asym2}:
\begin{eqnarray}
\gamma &=& u_n - \frac{\left(-1 + \frac{n -1}{n}\right)^{4}}{u_n -1} - \left(-1 + \frac{n -1}{n}\right)^{3} \left(2 + 2 \frac{n -1}{n}\right) \log{\left (- u_n + 1 \right )} \nonumber\\
       &+& \frac{\left(-4 + 2 \frac{n -1}{n}\right) \left(n -1\right)^{3} \log{\left (u_n \right )}}{n^{3}} - \frac{\left(n -1\right)^{4}}{n^{4} u_n} - 1\\
\alpha_1(k) &=& \frac{4 k^{3} - 6 k^{2} + 4 k -1}{n^{4}}  \\
\alpha_2(k) &=&  2 \frac{6 n k^{2} - 6 n k + 2 n - 4 k^{3} + 6 k^{2} - 4 k + 1}{n^{4}}  \\
\alpha_3(k) &=& - \frac{- \left(- n + k\right)^{4} + \left(n - k + 1\right)^{4}}{n^{4}} \\
\alpha_4(k) &=& 2 \frac{2 n^{3} - 6 n k^{2} + 6 n k - 2 n + 4 k^{3} - 6 k^{2} + 4 k -1}{n^{4}}       
\end{eqnarray}
\bigskip
\section{Variance of the Black-Karasinski process} 
\label{appB}
\begin{eqnarray}
\mathrm{Var}[X_{t}]&=&\mathbb{E}[(X_{t})^{2}] - \mathbb{E}[(X_{t})]^{2}\\ \nonumber
\mathbb{E}[(X_{t})^{2}] &=&  \mathrm{exp} \Bigl\{2\,\mathrm{ln}(X_{0}) e^{-k t} + 2a (1-e^{- k t}) + \frac{\sigma^{2}}{k}(1-e^{- 2k t}) \Bigr\}	\\ \nonumber
\mathbb{E}[(X_{t})]&=&  \mathrm{exp} \Bigl\{\mathrm{ln}(X_{0}) e^{-k t} + a (1-e^{- k t}) + \frac{\sigma^{2}}{4k}(1-e^{- 2k t}) \Bigr\}
\end{eqnarray}
\cleardoublepage

\end{document}